# The compliance of the molecular hydride superconductor BiH$_4$ with the Migdal's theorem

E. F. Talantsev, Yu. V. Blinova, and A. V. Korolev

*M. N. Miheev Institute of Metal Physics, Ural Branch, Russian Academy of Sciences, 18, S. Kovalevskoy St., Ekaterinburg 620108, Russia*

**Abstract**

The discovery of near-room-temperature superconductivity in H$_3$S sparked experimental and theoretical studies of highly compressed hydrides with the aim of obtaining room-temperature superconductivity. There are two dominant hydride classes where the search is ongoing: the first class is the covalently bonded hydrides (which is represented by H$_3$S and H$_3$P), and the second class is the clathrate-type hydrides (which is represented by LaH$_{10}$, YH$_6$, CaH$_6$, and other). In these classes, the hydrogen ions form three-dimensional sub-lattices involving the dissociation of hydrogen molecules (where, the H-H distance is 74 pm) into a metallic state with the shortest H-H distance of 151 pm in H$_3$S and 115 pm in LaH$_{10}$. Recently, the third class of highly compressed superconducting hydrides, where the hydrogen remains its molecular form, has been discovered. This class is represented by two superhydrides, BaH$_{12}$ and BiH$_4$, which both exhibit the smallest H-H distance of 81 pm in the pressure ranges where the highest $T_c$ is achieved. Here, we analyzed the available experimental data for the molecular hydrides BaH$_{12}$ and BiH$_4$. We found that BaH$_{12}$ exhibits nanosized grains of an average size of 26 nm and a low level of microstrain of 0.1%, both of which are independent of pressure in the range of $126\ GPa < P < 160\ GPa$. We also derived the Debye $\Theta_D$ and Einstein $\Theta_E$ temperatures, and the electron-phonon coupling constant $\lambda_{e-ph}$, in BaH$_{12}$ and BiH$_4$. The latter differs from the values obtained by first-principles calculations. The derived Fermi temperature $T_F \cong 20{,}000\ K$ for BiH$_4$ positions this molecular hydride between the unconventional and conventional superconductors bands in the Uemura plot. This position is outside of the band where covalently bonded and clathrate hydrides are located. The ratio $\Theta_D/T_F = 0.026$ of BiH4 is typical for pure metals and A-15 alloys. This implies that the BiH4, is the first hydride superconductor which complains with the Migdal's theorem.

**The compliance of the molecular hydride superconductor BiH₄ with the Migdal's theorem**

**I. Introduction.**

The discovery of a superconducting state in pressurized H₃S [1] with a transition temperature $T_c = 203\ K$ represented the triumph of modern experimental and theoretical physics. During the decade following this discovery, about six hundreds of new hydride phases were discovered during experimental quests and first-principles calculations [2–35]. The current progress in the field has been reviewed [36–41]. During the last decade, several important questions on fundamental properties of highly compressed and ambient pressure superconductors have been discussed [42–53].

There are two dominant classes of superhydrides that are primarily studied. The first class is the compounds with covalently bonded hydrogen. Historically, this class of high-$T_c$ hydrides was first discovered experimentally [54,55].

The second class is the clathrate hydrides, where there are several sign superconducting phases, for instance, LaH₁₀ [4–6,56], YH₆ [8,56,57], CaH₆ [58,59], and some ternary and quarterly hydrides.

However, both of these classes of hydrides share a common feature - a three-dimensional sublattice formed by dissociated hydrogen ions, which implies that the molecular form of hydrogen in these compounds has dissociated. While hydrogen molecules have an H–H distance of 74 pm, these three-dimensional metallic conducting sublattices in covalently bonded and clathrate hydrides exhibit much larger H–H distance (151 pm in H₃S and 115 pm in LaH₁₀. [60]).

It should be noted, that the hydrogen in metals is the topic for studies for more than 150 years [61], about a hundred years ago, it was found that atomic hydrogen does not react with pure bismuth [62]. Despite a fact that the electronegativity χ (and, thus, chemical properties) of the metals at high pressure are radically different from the ones at normal conditions [63] (for instance, elemental bismuth is changing its electronegativity from $\chi(P = 0\ GPa) = 1.2$ to $\chi(P = 200\ GPa) = -5.2$ [63]) we should not exclude that atomic hydrogen can occupy some interstitial sites in the crystal lattice at high pressures, similar to the occupation of interstitial sites in alloys [64] at ambient pressure. Hydrogen atoms vibrate mainly with localized (optical) modes at their interstitial sites [65], and, in addition, at the modes associated with strongly distorted (by the presence of the hydrogen) lattice vibrations (which are short-wavelength acoustic modes of the distorted lattice [65]). All mentioned above reasons cause the change in the phonon spectrum, and as a consequence of it, the electron-phonon coupling constant $\lambda_{e-ph}$ might also enhance.

From other hand, theoretical studies, including first-principles calculations (fpc), have predicted [66,67] that to achieve metallic conductivity and a superconducting state at liquid nitrogen temperatures in compressed hydrogen, the formation of a sublattice of dissociated molecular hydrogen is not strictly necessary.

And recently, a third class of highly compressed superconducting hydrides has been experimentally discovered, in which hydrogen retains its molecular form but exhibits metallization when compressed in the megabar range [68,69]. This class is currently represented by two superhydrides, $BaH_{12}$ and $BiH_4$, which both exhibit the smallest H-H distance of 81 pm in the pressure ranges [69], where the highest $T_c$ is achieved for each compound.

The first-principles study of the phase diagram and superconductivity in Bi-H system was performed by Ma *et al.* [70] in 2015. And to the best of our knowledge Shan *et al.*[69] performed first experimental study of this system.

Here we have analysed the currently available experimental data for the $BaH_{12}$ [68] and $BiH_4$ [69] molecular hydrides.

## II. Utilized models

There is no direct microscopy with submicron resolution, which can be used to reveal structural parameters of highly compressed materials in the diamond anvil cell (DAC). In attempt to resolve this problem, in [71,72] we proposed to use the Williamson-Hall (WH) analysis[73] of the X-ray diffraction (XRD) data:

$$\beta(\theta, P) = \frac{k_s \times \lambda_{X-ray}}{D(P) \times cos(\theta)} + 4 \times \varepsilon(P) \times tan(\theta), \quad (1)$$

where $\beta(\theta, P)$ is integral breadth of the peak in experimental XRD scan, $k_s$ is the Scherrer constant usually assigned as 0.9 [74–77], $\lambda_{X-ray}$ is the wavelength of the radiation, $D(P)$ is the mean size of coherent scattering regions, and the $\varepsilon(P)$ is the mean microstrain in crystalline lattice. Experimental XRD scans for $BaH_{12}$ hydride were kindly provided by the authors [68]. We processed these datasets by the DIOPTAS software [78].

One of the primary parameters of any electron-phonon superconductors is the Debye temperature $\Theta_D$ which can be deduced from the fit of measured thermal dependence of normal state resistance $R(T)$ to the Bloch- Grüneisen (BG) equation [79,80]:

$$R(T) = R_0 + A \left(\frac{T}{\Theta_D}\right)^5 \int_0^{\frac{\Theta_D}{T}} \frac{x^5}{(e^x-1)(1-e^{-x})} dx \quad (2)$$

where where $R_0$, $A$, and $\Theta_D$ are free-fitting parameters. Deduced $\Theta_D$ and measured in experiment $T_c$ can be used to deduce the electron-phonon coupling constant $\lambda_{e-ph}$ as a root of the system of McMillan equations [81,82] :

$$T_c = \frac{\Theta_D}{1.4} e^{\left[-\frac{1.04(1+\lambda_{e-ph})}{\lambda_{e-ph}-\mu^*(1+0.62\lambda_{e-ph})}\right]} \cdot f_1 \cdot f_2^* \quad (3)$$

$$f_1 = \left(1 + \left(\frac{\lambda_{e-ph}}{2.46(1+3.8\mu^*)}\right)^{3/2}\right)^{1/3} \quad (4)$$

$$f_2^* = 1 + (0.0241 - 0.0735 \cdot \mu^*) \cdot \lambda_{e-ph}^2 \quad (5)$$

where $\mu^*$ is the Coulomb pseudopotential ($\mu^* = 0.10 - 0.16$ [7,9]) for which we used the mean value of $\mu^* = 0.13$ below. Derived $\lambda_{e-ph}$ can be compared with the value calculated by the first principles calculations.

Recently, Pinsook and Tanthum [83] proposed that experimental $R(T)$ can be fitted to the equation based on the Einstein temperature, $\Theta_E$:

$$R(T) = R_0 + A \cdot \frac{\frac{\Theta_E^2}{T}}{\left(e^{\frac{\Theta_E}{T}}-1\right)\left(1-e^{-\frac{\Theta_E}{T}}\right)} \tag{6}$$

where $R_0$, $A$, and $\Theta_E$ are free-fitting parameters. Pinsook and Tanthum [83] found that the deduced $\Theta_E$ is approximately equal to the temperature $\left(\frac{\hbar}{k_B}\omega_{ln}\right)$ associated with the logarithmically averaged characteristic frequency $\omega_{ln}$ in the Allen-Dynes approximation [84,85] of the Eliashberg theory of electron-phonon superconductivity [86]:

$$T_c = \frac{1}{1.20}\left(\frac{\hbar}{k_B}\omega_{ln}\right) e^{\left[-\frac{1.04(1+\lambda_{e-ph})}{\lambda_{e-ph}-\mu^*(1+0.62\lambda_{e-ph})}\right]} \cdot f_1 \cdot f_2 \tag{7}$$

$$f_1 = \left(1 + \left(\frac{\lambda_{e-ph}}{2.46(1+3.8\mu^*)}\right)^{3/2}\right)^{1/3} \tag{8}$$

$$f_2 = 1 + \frac{\left(\frac{\langle\omega^2\rangle^{1/2}}{\omega_{ln}}-1\right)\lambda_{e-ph}^2}{\lambda_{e-ph}^2 + \left(1.82(1+6.3\mu^*)\left(\frac{\langle\omega^2\rangle^{1/2}}{\omega_{ln}}\right)\right)^2} \tag{9}$$

$$\langle\omega^2\rangle^{1/2} = \frac{2}{\lambda_{e-ph}}\int_0^\infty \omega \times \alpha^2 \times F(\omega)d\omega \tag{10}$$

In [82,87] we proposed to replace Eqs. 8,9 by approximated Eq. 5, because the values of $\alpha^2 \times F(\omega)$ and $\omega_{ln}$ can not be deduced from experimental $R(T)$ datasets. Thus, we also used the following equation:

$$T_c = \frac{\Theta_E}{1.20} e^{\left[-\frac{1.04(1+\lambda_{e-ph,E})}{\lambda_{e-ph}-\mu^*(1+0.62\lambda_{e-ph,E})}\right]} \cdot f_1 \cdot f_2^* \tag{11}$$

for the independent determination of the electron-phonon coupling constant, $\lambda_{e-ph,E}$, from the deduced Einstein temperature, $\Theta_E$.

In this work we derived the ground state superconducting coherence length $\xi(0)$ from the fit of the upper critical field data $B_{c2}(T)$ dataset to the single-band analytical approximation[88] for the Helfand-Werthamer theory [89]:

$$B_{c2}(T) = \frac{\phi_0}{2\pi\xi^2(0)} \times \left( \frac{1-\left(\frac{T}{T_c}\right)^2}{1+0.42\times\left(\frac{T}{T_c}\right)^{1.47}} \right) \qquad (12)$$

where $\phi_0$ is the superconducting flux quantum.

### III. Parameters of the BaH$_{12}$ phase

To determine the instrumental broadening of the XRD machine used in [68], we analyzed XRD scan for calibrated $CeO_2$, which was kindly provided to us by D. Semenok [68]. Data for $CeO_2$ was procced by DIOPTAS [78] software. XRD peaks were fitted to the sum of gaussians:

$$I(2\theta) = A_{bg} + \sum_{k=1}^{N} \frac{A_k}{w\sqrt{\frac{\pi}{2}}} \times e^{-2\frac{\left(2\theta-2\theta_{peak,k}\right)^2}{w^2}}, \qquad (13)$$

where $A_k$ is the peak area, $2\theta_{peak,k}$ is the peak position, $\beta_k(2\theta) = w\sqrt{\frac{\pi}{2}}$ is peak integral breadth, and $A_k$, $2\theta_{peak,k}$, and $w$ are-free fitting parameters. The background level, $A_{bg}$, was a manually adjusted parameter in each XRD fit in the DIOPTAS software [78] and it was a free fitting parameter in final fit in Origin software [90].

Derived peaks breadth, $\beta_i(2\theta)$, were fitted to Caglioti equation [91]:

$$\beta_{inst}(2\theta) = \sqrt{U\cdot tan^2\left(\frac{2\theta}{2}\right) + V\cdot tan\left(\frac{2\theta}{2}\right) + W}, \qquad (14)$$

where $U, V$, and $W$ are free-fitting parameters. Result of the fit is shown in Fig. 1, where derived parameters are: $U = 0, V = 0, W = (1.29 \pm 0.06) \times 10^{-6}$.

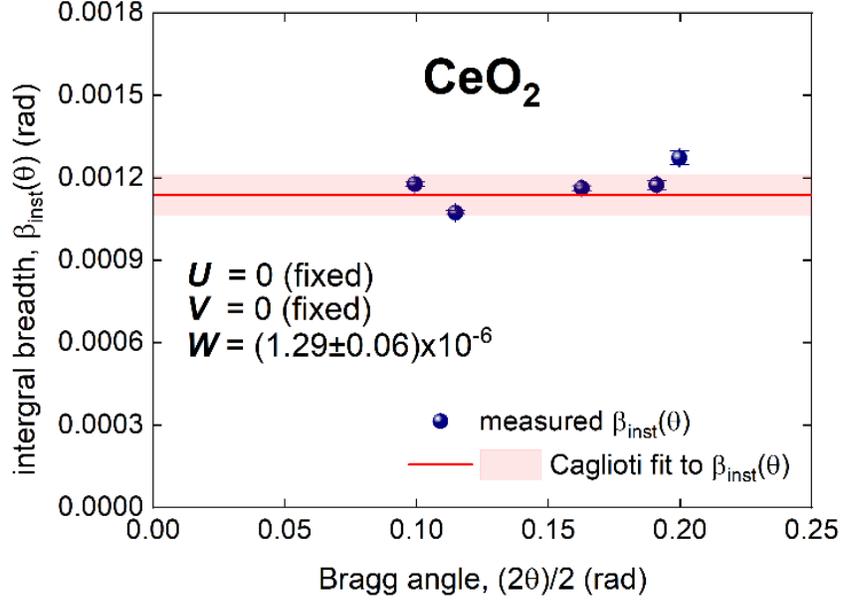

**Figure 1.** Experimental $\beta_i(\theta)$ data for standard $CeO_2$ sample. Raw XRD scan was provided by D. Semenok. 95% confidence bands are shown by pink shadow area.

Peaks in XRD scans of $BaH_{12}$ were fitted to Eq. 13 and $\beta_{measured}(2\theta)$ datasets were deduced. The peaks breadth originated from the $BaH_{12}$ phase, $\beta_{BaH12}(2\theta)$, were calculated by the equation:

$$\beta_{BaH12}(2\theta) = \sqrt{\beta_{measured}^2(2\theta) - \beta_{inst}^2(2\theta)}, \tag{15}$$

Derived $\beta_{BaH12}(\theta, P)$ datasets were fitted to the WH equation (Eq. 1). Deduced values from these fits, $D(P)$ and $\varepsilon(P)$, are shown in Fig. 2.

One can see (Fig. 2) that $D(P)$ and $\varepsilon(P)$ are practically independent from applied pressure in the studied range of $126\ GPa \leq P \leq 160\ GPa$. It is interesting to note, that highly compressed $BaH_{12}$ has negligible level of micro-strain.

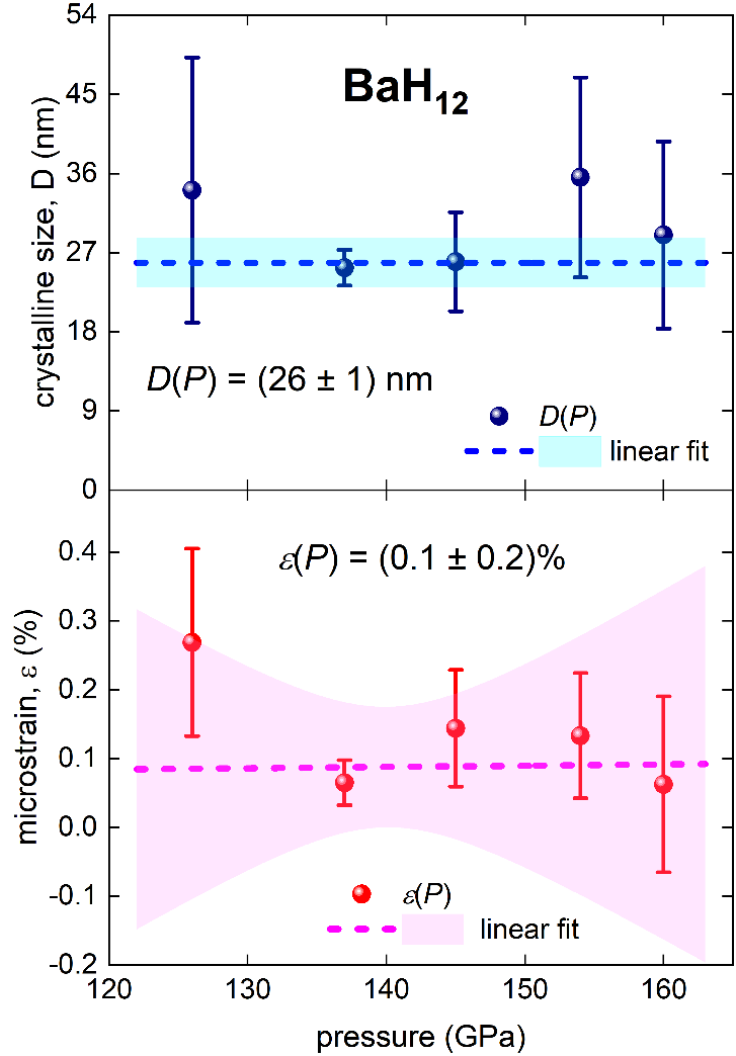

**Figure 2.** Derived (a) grain size $D(P)$ and (b) micro-strain $\varepsilon(P)$ in highly compressed BaH$_{12}$. Raw XRD data was measured by Chen *et al.* [68]. 95% confidence bands are shown by (a) cyan and (b) magenta shadow areas.

In Figures 3 and 4, we showed the $R(T, P = 100\ GPa)$ data measured in BaH$_{12}$ by Chen *et al.* [68] and data fits to Bloch-Grüneisen (Eq. 2) and Einstein (Eq. 6) models in panels ***a*** and ***b***, respectively. The substitution derived $\Theta_D$ and $\Theta_E$ in respectful equations returns the $\lambda_{e-ph,D}$ and $\lambda_{e-ph,E}$ values which appear to be practical equal to each other $\lambda_{e-ph,E} = \lambda_{e-ph,D} \cong 0.71$ (Fig. 3), and $\lambda_{e-ph,E} = 0.70 \cong \lambda_{e-ph,D} = 0.69$ (Fig. 4). We also found that $\Theta_E/\Theta_D \cong 0.805$ (Figs. 3,4), which is lower than the theoretical value of $\Theta_E/\Theta_D = 0.828$. It should be noted that the obtained $\Theta_E/\Theta_D \cong 0.805$ value is very close to the one deduced for highly compressed LaB$_2$H$_8$[87].

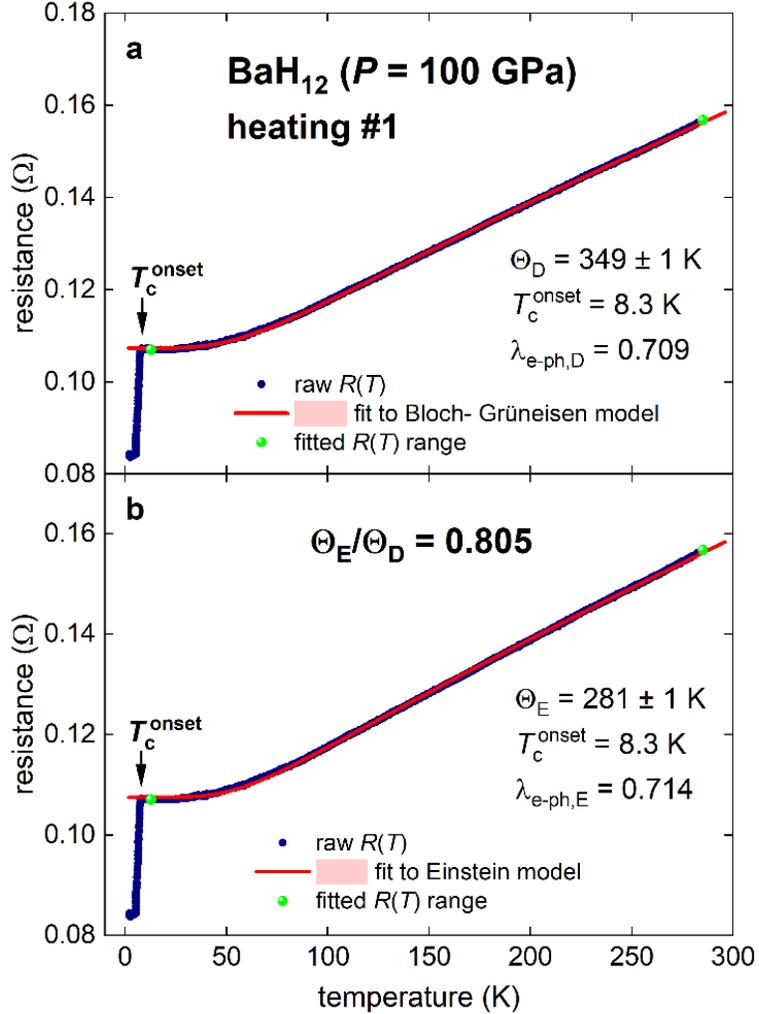

**Figure 3.** $R(T, P = 100\ GPa)$ dataset for BaH$_{12}$ (heating #1) and data fit to (**a**) Bloch-Grüneisen (Eq. 2) and (**b**) Einstein (Eq. 6) models. Raw $R(T, P = 100\ GPa)$ data reported by Chen *et al.* [68]. Derived parameters are shown in panels (**a**) and (**b**). Goodness of fit: (**a**) $R = 0.9998$ and (**b**) $R = 0.9997$. 95% confidence bands are shown by pink shadow areas.

Moreover, the derived values of $\lambda_{e-ph}(P = 100\ GPa) \cong 0.70$ differs from the value $\lambda_{e-ph}(P = 140\ GPa) = 0.95$ obtained by first principles calculations [68]. Such a difference can originate from the difference in pressure used in experiment ($P = 100\ GPa$) and the one used for calculations ($P = 140\ GPa$). Another probable reason is the presence of vacancies in the crystalline structure of BaH$_{12}$. Atomic-scale vacancies are now considered an important structural defect that can determine $T_c$ in high-pressure superconductors [35,92].

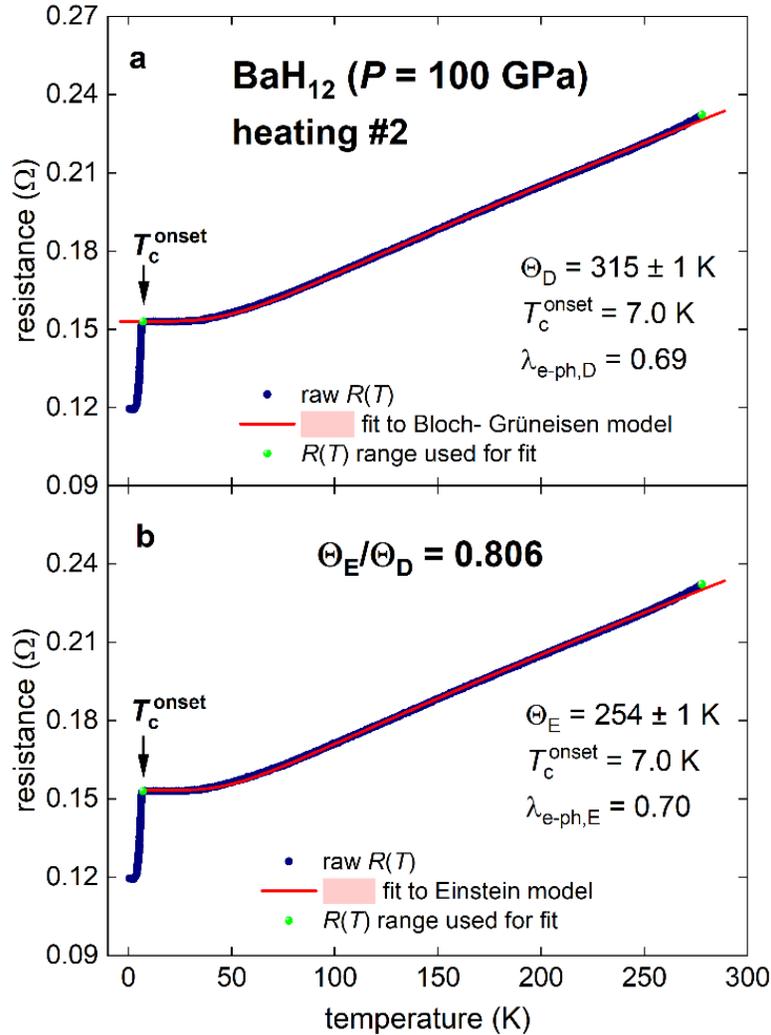

**Figure 4.** $R(T, P = 100\ GPa)$ dataset for BaH$_{12}$ (heating #2) and data fit to (**a**) Bloch-Grüneisen (Eq. 2) and (**b**) Einstein (Eq. 6) models. Raw $R(T, P = 100\ GPa)$ data reported by Chen *et al.* [68]. Derived parameters are shown in panels (**a**) and (**b**). Goodness of fit: (**a**) $R = 0.9998$ and (**b**) $R = 0.9997$. 95% confidence bands are shown by pink shadow areas.

**IV. Parameters of the BiH$_4$ phase**

In Figure 5, we showed the $R(T, P = 172\ GPa)$ data measured in BiH$_4$ hydride by Shan *et al.*[69]. Data fits to Bloch-Grüneisen (Eq. 2) and Einstein (Eq. 6) models are shown in panels *a* and *b* of Fig. 5, respectively.

To calculate the $\lambda_{e-ph,D}$ and $\lambda_{e-ph,E}$ values we defined the transition temperature by a strict resistance criterion $\frac{R(T=T_c)}{R_{norm}} \to 0.05$. The substitution of the derived $\Theta_D$ and $\Theta_E$ in respectful equations returns the $\lambda_{e-ph,D}$ and $\lambda_{e-ph,E}$ values which appear to be equal to each other, $\lambda_{e-ph,E} = \lambda_{e-ph,D} = 2.9$ (Fig. 5), In addition, we found that the ratio $\Theta_E/\Theta_D = 0.822$ (Fig. 5) is practically equal to the theoretical value of $\Theta_E/\Theta_D = 0.828$.

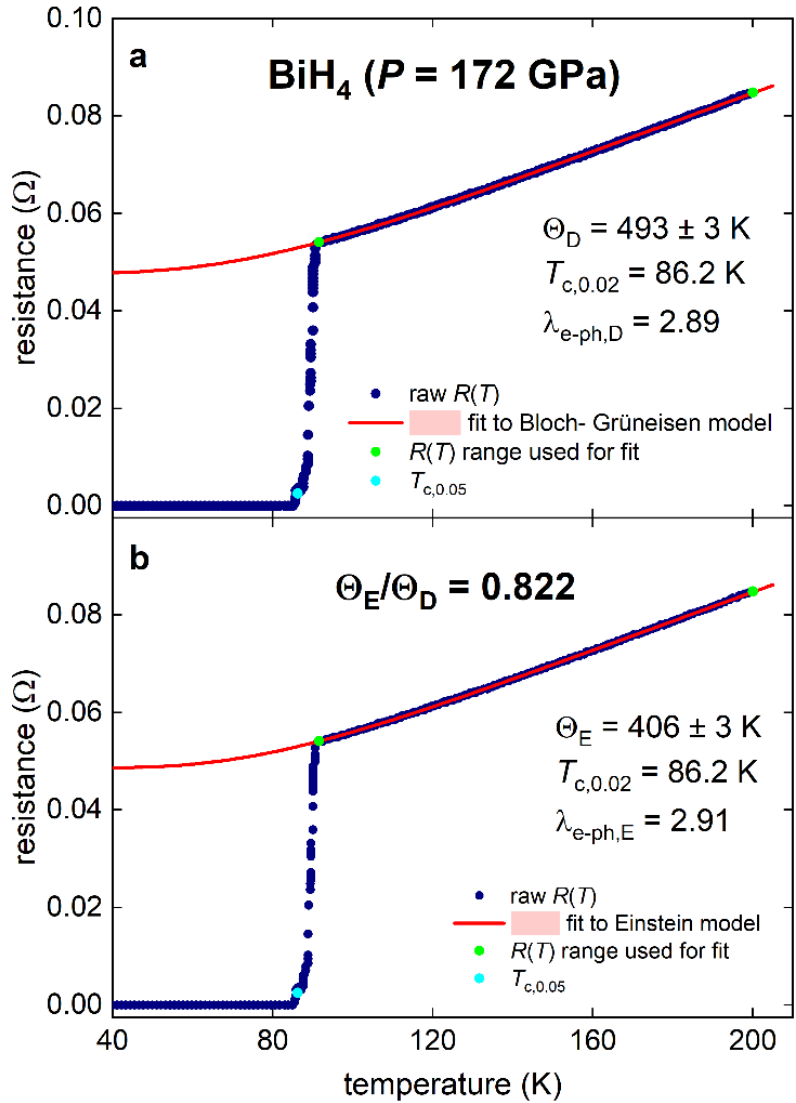

**Figure 5.** $R(T, P = 172\ GPa)$ dataset for BiH$_4$ and data fit to (**a**) Bloch-Grüneisen (Eq. 2) and (**b**) Einstein (Eq. 6) models. Raw $R(T, P = 100\ GPa)$ data reported by Shan *et al.* [69]. Derived parameters are shown in panels. Goodness of fit is $R = 0.9998$ for both panels. 95% confidence bands are shown by pink shadow areas.

The derived $\Theta_E/\Theta_D$ values for BaH$_{12}$ and BiH$_4$ confirm the proposal by Pinsook and Tanthum[83] that the Einstein temperature $\Theta_E$ can be used instead of the value $\left(\frac{\hbar}{k_B}\omega_{ln}\right)$ in the Allen-Dynes equation[85].

Considering results of the first-principles calculations by Ma *et al.* [70], we found that the closest theoretical $\left(\frac{\hbar}{k_B}\omega_{ln}\right)$ value to our value $\Theta_E(P = 172\ GPa) = 406\ K$ (deduced from experimental $R(T)$) is $\left(\frac{\hbar}{k_B}\omega_{ln}\right)(P = 150\ GPa) = 507\ K$ calculated for *P21/m*-BiH$_2$ phase.

Moreover, the deduced $\lambda_{e-ph,E}(172\ GPa) = \lambda_{e-ph,D}(172\ GPa) = 2.9$ is remarkably different from the value calculated by first-principles calculations $\lambda_{e-ph,fpc}(170\ GPa) = 2.01$ [69] and $\lambda_{e-ph,fpc}(150\ GPa) \leq 1.34$ for BiH$_n$ (*n* = 2,3,4,5,6).

Such a large difference between the deduced $\lambda_{e-ph}$ from experimental $R(T)$ (by utilizing two approaches Eqs. 2-11) and the calculated values by first-principles has not been reported before [83,93]. Instead, it was shown a reasonable agreement between experimental and calculated values [83,93]. Based on this, we can state that the deduced value of $\lambda_{e-ph}(172\ GPa) = 2.9$ is correct.

Such a large difference should be commented. Our current interpretation of this difference is based on idea that some small part of the molecular hydrogen is actually dissociated at such high pressure of $P = 172\ GPa$. The possibility of partial dissociation of molecular hydrogen cannot be excluded, based on available literature which we reviewed in the *Introduction* part. If so, the dissociated hydrogen makes contribution in the phonon spectrum. This contribution distorts the phonon spectrum, and, perhaps, creates new bands in addition to the bands calculated to the date [69,70] in BiH4. In the result, the $\lambda_{e-ph}$ value can be enhanced in comparison with the $\lambda_{e-ph,fpc}$ value, which is calculated based on the assumption that the real structure of the BiH$_4$ is formed by bismuth and molecular hydrogen only[69,70].

In Figure 6, we showed the upper critical field $B_{c2}(T)$ which was deduced from the $R(T, B, P = 172\ GPa)$ data reported by Shan et al.[69]. For the derivation, we used the resistive criterion of $\frac{R(T=T_c)}{R_{norm}} \to 0.07$.

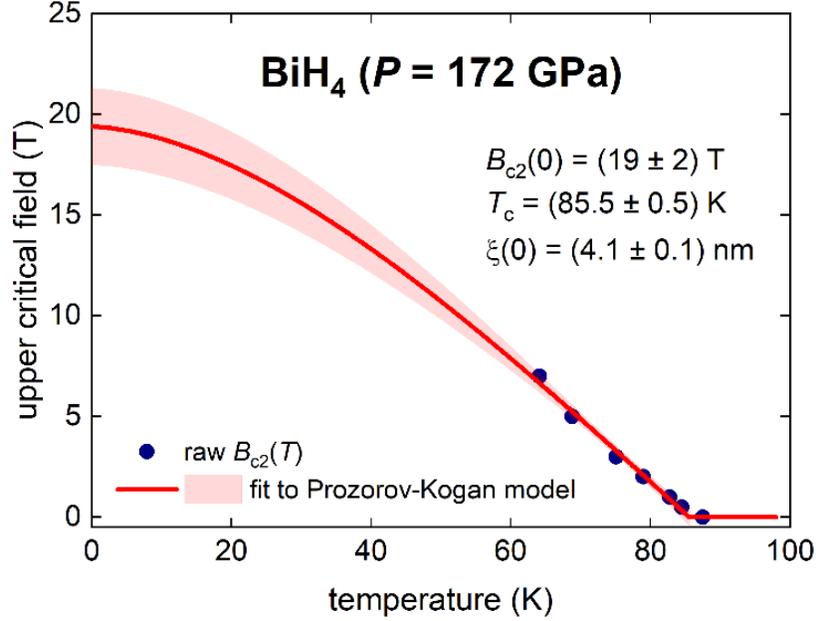

**Figure 6.** $B_{c2}(T)$ for BiH$_4$ and data fit Eq. 12. Raw $R(T, B, P = 172\ GPa)$ data reported by Shan et al.[69] $B_{c2}(T)$ datasets were derived by applying criterion $\frac{R(T=T_c)}{R_{norm}} = 0.07$. 95% confidence intervals are indicated by the pink shaded area.

The ground state coherence length, $\xi(0) = 4.1 \pm 0.1\ nm$ was deduced from the $B_{c2}(T)$ fit to the Equation 12 (Figure 6). The deduced value was substituted in the Equation 11 to calculate the Fermi temperature:

$$T_F = \frac{\pi^2 m_e}{8 \cdot k_B} \cdot (1 + \lambda_{e-ph}) \cdot \xi^2(0) \cdot \left(\frac{\alpha k_B T_c}{\hbar}\right)^2, \tag{16}$$

where $m_e$ is bare electron mass, $\hbar$ is reduced Planck constant, $\alpha = \frac{2\Delta(0)}{k_B T_c}$, and $\lambda_{e-ph} = 2.9$ was taken from Fig. 5. The $\frac{2\Delta(0)}{k_B T_c}$ ratio was calculated by linear approximated relation proposed in Ref.[94]:

$$\alpha \equiv \frac{2\Delta(0)}{k_B T_c} = 3.26 + 0.74 \times \lambda_{e-ph}. \qquad (17)$$

Calculated $T_F = 19{,}500\ K$ implies that the BiH$_4$ falls between the unconventional and conventional superconductors bands in the Uemura plot with $T_c/T_F = 0.0044$ (Fig. 7). This position is outside (towards the BCS superconductivity band) the band where covalently bonded and clathrate hydrides are located (Fig. 7).

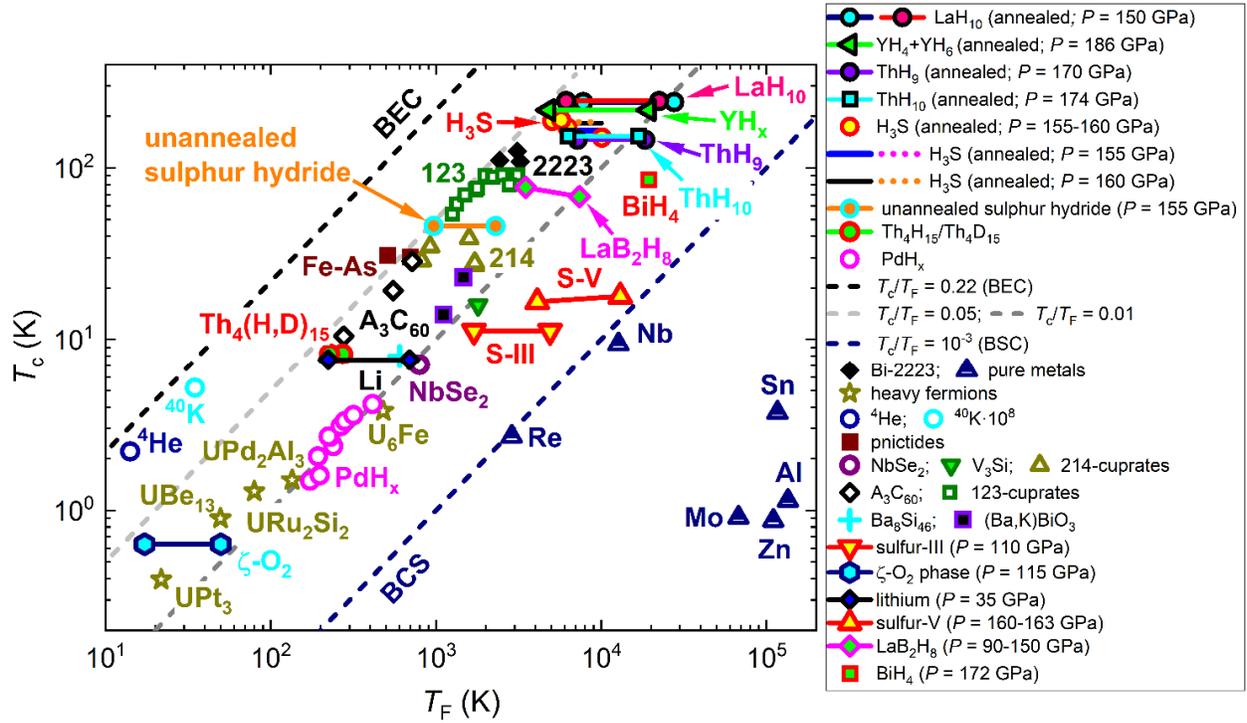

**Figure 7.** Uemura plot ($T_c$ vs. $T_F$) where BiH$_4$ phase is shown together with main superconducting families. References to the original data are in Refs.[72,95–97]. The position of the BiH$_4$ phase is outside of the band, where other hydrogen-based superconductors are located, and toward the band of classical BCS superconductors.

However, the most important finding in this work is the established ratio of $\Theta_D/T_F = 0.026$ in BiH$_4$ (Fig. 8). This value is similar to the $\Theta_D/T_F$ in pure metals and A-15 alloys (Fig. 8). Thus, the BiH$_4$ is the first discovered highly compressed high-$T_c$ superhydride [72,93,95,96] which complies [98–103] with the Migdal's theorem [104]. This is unexpected result. Perhaps this result is additional evidence for the presence of the dissociated hydrogen in the structure of highly compressed BiH4, because the metallization of the dissociated hydrogen should bust the Fermi surface size.

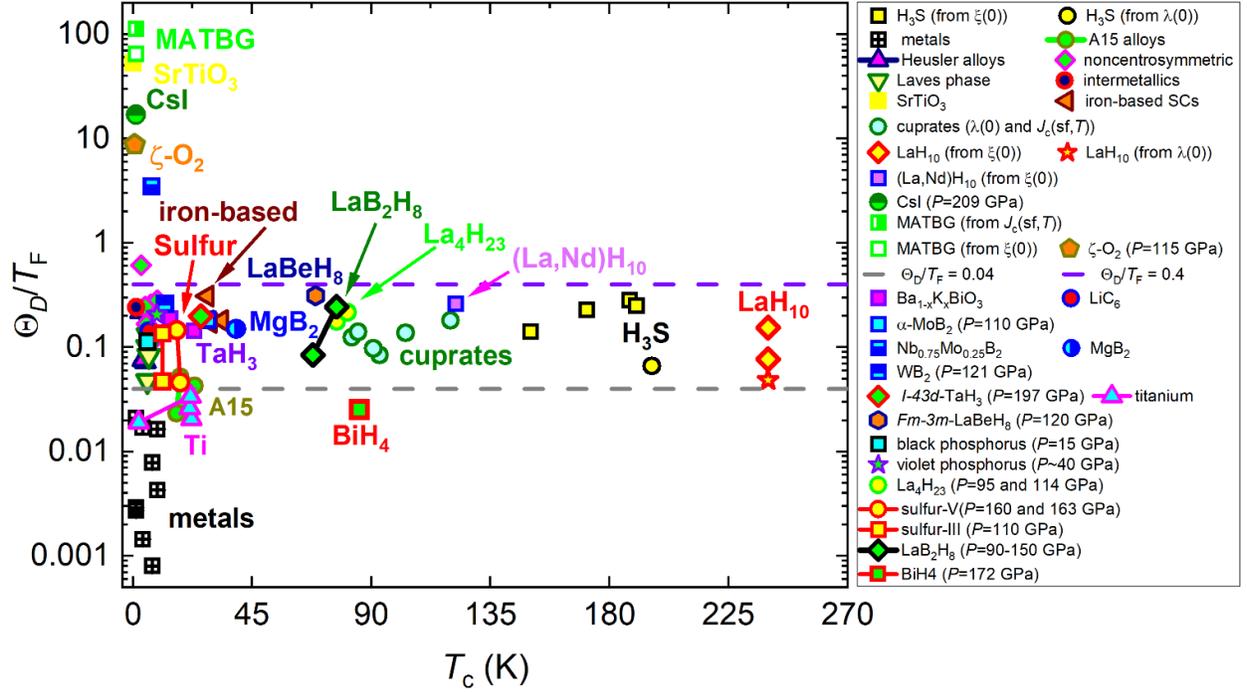

**Figure 7.** $\frac{\Theta_D}{T_F}$ vs $T_c$ plot where BiH4 phase is shown together with main superconducting families. References to the original data are in Refs.[72,93,95,96].

## V. Conclusions

In this study, we analysed available experimental data reported for highly compressed molecular hydrides BaH$_{12}$ [68] and BiH$_4$ [69].

In the result, we confirm that the proposal by Pinsook and Tanthum [83] that the Einstein temperature $\Theta_E$ can be used instead of the $\left(\frac{\hbar}{k_B}\omega_{ln}\right)$ term in the Allen-Dynes equation [85].

One of the central findings of our study is that the BiH$_4$ exhibits a low level of nonadiabaticity, $\Theta_D/T_F = 0.026$, and, thus, the BiH$_4$ is the first hydride which complies with Migdal theorem [104].

We have proposed that the BiH$_4$ samples at $P = 172\ GPa$ contain some hydrogen in dissociated metallized state.

**Acknowledgements**

The authors thanks Dr. Dmitrii Semenok (Center for High Pressure Science & Technology) for providing raw experimental data for molecular hydride BaH$_{12}$ [68] and for consultations in regard of DIOPTAS software. The work was carried out within the framework of the state assignment of the Ministry of Science and Higher Education of the Russian Federation for the IMP UB RAS.